\documentclass[iop]{emulateapj}
\usepackage{natbib}
\usepackage{amsmath}
\DeclareMathOperator{\std}{std}

\shorttitle{Near-surface Flows Around Active Regions and Flares}
\shortauthors{Braun}

\slugcomment{accepted Astrophys.J. Jan 28, 2016}

\begin{document}

\title{A Helioseismic Survey of Near-surface Flows Around Active Regions and their Association with Flares}

\author{D.~C.\ Braun}
\affil{NorthWest Research Associates, 3380 Mitchell Lane, 
Boulder, CO 80301, USA}
\email{dbraun@cora.nwra.com}

\begin{abstract}
We use helioseismic holography to study the association of shallow flows
with solar flare activity in about 250 large sunspot groups observed
between 2010 and 2014 with the Helioseismic and Magnetic Imager on the Solar Dynamics
Observatory. Four basic flow parameters: horizontal speed, 
horizontal component of divergence, vertical component of 
vorticity, and a vertical kinetic helicity proxy, are mapped for each
active region during its passage across the solar disk. Flow indices are derived representing
the mean and standard deviation of these parameters over magnetic masks and
compared with contemporary measures of flare X-ray flux. A correlation exists for
several of the flow indices, especially those based on the speed and the standard deviation
of all flow parameters. However, their correlation with X-ray flux is similar to that 
observed with the mean unsigned magnetic flux density over the same masks. 
The temporal variation of the flow indices are studied, and a superposed epoch  
analysis with respect to the occurrence to 70 M and X-class flares is made.
While flows evolve with the passage of the active regions across the disk, no
discernible precursors or other temporal changes specifically associated with flares are
detected. 

\end{abstract}


\section{Introduction}\label{intro}

A fundamental question in solar physics is
how magnetic fields emerge from the solar convection zone into the solar atmosphere 
and then develop solar eruptions such as flares and CMEs 
\citep[][]{Fan2009, Schrijver2009}. 
It is widely held that highly twisted magnetic fields, emerging perhaps 
into a pre-existing
field, are required for the formation of active regions producing M- or X-class 
flares \citep{Schrijver2009}.
Observations of the linkage between 
subsurface flows and the twist, electric current content, 
and other properties of active regions (hereafter ARs) 
are important to understand, and perhaps even predict, the flare phenomenon.
Considerable effort has been spent in studying the time evolution of
magnetograms or continuum images of active regions in order to quantify 
photospheric motions which may lead to predictive indices 
\citep[e.g.][]{Leka2007, Schrijver2007, Welsch2009}. 
Helioseismic measurements of subsurface motions extend the spatial volume over which
AR properties, and their role in transporting magnetic helicity and current into the corona,
can be explored. In addition, the direct inference of mass flows 
complements measurements of photospheric motions of (magnetic or continuum) 
features \citep{Wang2011,Beauregard2012,Wang2014}.

Any detection and interpretation of photospheric or subsurface flow precursors
to flares requires understanding the pre-existing flow fields, and their general
characteristics,  associated with
active regions. Photospheric outflows (called ``moats'') extending past the penumbrae of 
most sunspots have
been studied for decades \citep[][]{Brickhouse1988}. Larger converging flows
around active regions were first detected using local helioseismology 
\citep[][]{Gizon2001,Haber2001}.  Inverse modeling using ring-diagram (hereafter
RD) analysis performed by \cite{Haber2004} indicated these converging flows are situated
above deeper outflows. A consensus appears to be that these flows have
speeds on the order of 50 m s$^{-1}$  which may extend
out to as much as 30$^\circ$ from the AR centers. However,
\cite{Braun2011a} found that most outflows 
around ARs are compact and have flow speeds typical of the surrounding 
supergranulation.

\cite{Komm2004} first began to systematically characterize AR flows, deduced
from low-resolution RD analysis, in terms of their horizontal divergence and
their vertical components of vorticity and kinetic helicity. That work was the
first to detect a hemispheric preference for the vertical 
vorticity of the flows, which was cyclonic over the 15$^\circ$ spatially-averaged
flows both with and without the removal of a large-scale
differential rotation pattern.
Using ring-diagram procedures with considerably higher resolution,
\cite{Hindman2009} examined the divergence and
vortical components of motions within about 200 magnetic regions.
They confirmed cyclonic motions near active region boundaries, but also demonstrated
an anticyclonic trend associated with the cores of the ARs which are presumably
dominated by the sunspot moats. More recently, \cite{Komm2015c} have compared
the hemispheric preference, for long-lived activity complexes, of both 
the subsurface kinetic helicity, determined from ground-based observations from
the Global Oscillation Network Group (GONG), and the 
current helicity as determined
from synoptic vector magnetograms.

Specifically to examine the relation between subsurface flows, deduced from
from (low-resolution) RD analysis, and solar flares, \cite{Mason2006} surveyed 
the subsurface vorticity of ARs over 43 Carrington rotations using GONG 
and over 20 rotations observed using data from 
the Michelson Doppler Imager (MDI) instrument on
the SOHO spacecraft. Both data exhibited a trend between 
the unsigned vorticity and the product of the logarithm of flare intensity  
and maximum unsigned magnetic flux, for ARs above a given flux threshold.
An expanded version of the GONG RD-based survey was used by \cite{Komm2009}
to demonstrate a correlation between X-ray flux and vorticity, with both
quantities averaged over the disk passage of the ARs.
The flow measurements from this survey subsequently 
provided the basis for the development of an empirically-based 
parameter, based on the subsurface kinetic helicity density,
which showed specific temporal variations 2-3 days before flares 
\citep[including C, M, and X-class events;][]{Reinard2010}. 
Applying discriminant analysis to a variety of both magnetic and 
subsurface-flow parameters, \cite{Komm2011} suggest that the subsurface flow
parameters improve the ability to distinguish between flaring and non-flaring
ARs.  

Using a much smaller sample of 5 flaring ARs, 
but employing time-distance
helioseismic methods with considerably higher spatial resolution, 
\cite{Gao2014} found sporadic and short-duration
changes (called ``bumps'') in the kinetic helicity which, 
slightly more than half the time, occurred within about 8 hours (before
or after) an X-class flare. 

The primary goal in this study is to examine the association with solar flares
of near-surface flows within solar magnetic regions. This association may include:
(1) the predilection of flares to occur in regions with specific flow
properties, and (2) any precursor of, or response to, specific solar flares
visible in the temporal evolution of the flow patterns. To do this we make use
of nearly five years of nearly continuous and high resolution SDO/HMI observations of
approximately 250 of the largest sunspot groups cataloged by NOAA. Our motivation
is not to reproduce or confirm any prior study, but to exploit the high-resolution
capabilities of helioseismic holography to perform an independent examination
of the flows within the largest active regions. It is
necessary to place any observed 
association (or lack thereof) between flows and flares in the proper
context. Consequently, a sub-goal is to provide a survey of the general near-surface
flow properties of large NOAA sunspot groups. For this study, we focus our
attention on specific flow parameters which can be derived from inferences
of the two-dimensional near-surface vector flow field, including the horizontal
component of the divergence and the vertical component of the vorticity. We
also examine the product of these, which provides a proxy for the vertical
component of kinetic helicity \citep{Rudiger1999,Komm2007}. A fourth parameter
is the speed of the horizontal flow. Averages and 
standard-deviations of these basic parameters,
assessed over spatial masks constructed using HMI 
magnetograms, provide a set
of indices with with we search, using scatter plots and superposed epoch analysis, 
for correlations with solar flare activity.

\section{Active Region Selection}\label{ar_selection}

The sample of active regions we consider consists of all 
sunspot groups assigned a number by NOAA between the start of HMI
observations (2010 May) and 2014 December, and which reached a
size of at least 200 micro-hemispheres (hereafter $\mu$H). Although complete records of
the the sunspot group properties are 
available\footnote{\url{http:ngdc.noaa.gov/stp/space-weather/solar-data/solar-features/sunspot-regions/usaf\_mwl}}, 
for convenience we used the sunspot database maintained at NASA Marshall
Space Flight 
Center\footnote{\url{http:solarscience.msfc.nasa.gov/greenwch.shtml}}
which provides daily averages of the sunspot group properties including size
and Carrington coordinates.

We identified 252 regions which met these criteria, which represent approximately 
the largest 20\% of all NOAA numbered sunspot groups. 
Our survey includes a single region 
(AR 12017) which had a maximum area of only 160 $\mu$H but
was responsible for 23 flares including an X-class flare. Our final
sample includes ARs responsible for all (43) X-class flares, 
86\% ($\approx$460) of the M-class flares, and 72\% ($\approx$3600) C-class flares occurring
during this time period. 

In spite of the size criteria for selection, the sample includes  
cases of relatively quiescent ARs which provide a basis for comparison
with the more flare productive ones. For example, 137 ARs in the sample 
(i.e. slightly more than half) 
produced neither
M nor X-class flares and 17 of those produced no C-class flares either.
Unfortunately for the statistician, the Sun does not appear to emerge large numbers
of flare-free magnetic
regions with a distribution of properties (e.g. size, sunspot number, 
magnetic flux) which otherwise match the flaring regions. There are likely 
diminishing (and even detrimental) returns in including smaller flare-free
regions in our survey. That being said, we need to be mindful of selection
bias when (1) assigning meaning (e.g. cause-and-effect) to associations
of flows and flares, and (2) generalizing observations or inferences 
about the flow properties of our sample to other types of active regions.

For each AR, a 9.1 day long datacube is constructed from a Postel's projection
of the full-disk HMI Dopplergrams and
centered on the Carrington coordinates averaged
over its disk passage. The remapped datacube spans 30$^{\circ}$ 
by 30$^{\circ}$ with a pixel
spacing of 0.0573$^{\circ}$.
and is divided into 16 non-overlapping intervals of 13.6 hr duration for
helioseismic analysis. For context, a set of remapped, cospatial and 
time-averaged line-of-sight HMI magnetograms for each interval is constructed,
using full-disk magnetograms sampled every 68 minutes. 
Time intervals for which the mean AR location was greater than 60$^{\circ}$
from disk center, or for which gaps in the HMI data exceeded 30\% of the 13.6 hr period
are excluded from the survey. We are left with
3908 sets of helioseismic measurements, defining a set to be a single AR observed
over a unique 13.6 hr interval.

\section{Helioseismic Holography}\label{hh}

Helioseismic holography (hereafter HH) is a method which
computationally extrapolates the surface acoustic field
from a selected area or ``pupil'' into the solar interior \citep{Lindsey1997}
in order to estimate the amplitudes of the waves propagating into
or out of a focus point at a chosen depth and position in the solar
interior.  These amplitudes are called the acoustic ingression and
egression respectively. 
To study travel-time anomalies sensitive to the flows one 
constructs cross covariances between the ingression and egression
amplitudes using pupils which take the form of an annulus divided into 
quadrants.  This type of
pupil configuration is the basis for ``lateral vantage'' HH 
\citep{Lindsey2004b}, to which 
deep-focus methods in time-distance helioseismology
and common-depth-point reflection seismology methods are analogous.
For the results presented here, we choose a focus depth 3 Mm below the surface.

Travel-time measurements sensitive to horizontal flows are extracted from 
cross-covariances between the egressions and ingressions computed in
pupils spanning opposite quadrants which extend in the
east, west, north and south directions
from the focus.  The methodology is described in detail in prior publications 
\citep[e.g.][]{Braun2007,Braun2008b,Braun2014}. The steps include: (1)  
compute the 3D Fourier transform 
of the Postel-projected data 
in both spatial
dimensions and in time, (2) extract the data
within the frequency bandpass 2.5 -- 5.5 mHz,
(3) apply a phase-speed filter, (4) compute egression and ingression 
amplitudes with the appropriate Green's functions, (4) compute
egression--ingression correlations, and (5) measure travel-time differences.
The filter employed in step (3) helps to reduce noise
with high spatial-frequency. It consists of
a Gaussian function of phase speed with a full width at half maximum of 9.2 km s$^{-1}$ and
is centered at 18.8 km s$^{-1}$ which is tuned to waves
propagating horizontally at a depth of 3 Mm. 
The design and utility of
filters for lateral vantage HH is discussed further by \citet{Braun2014}.
Step (4) is computed by convolutions of the data cube with Green's functions
computed with the eikonal approximation and using a plane-parallel
approximation, which is well suited for the shallow focus depth of 3 Mm.
\cite{Braun2014} explores the validity of this approximation in detail.
The products of these analyses are maps of travel-time
shifts $\delta \tau_{ns}$, and $\delta \tau_{we}$, which represent
standard north-south and west-east travel-time differences.

\subsection{Flow calibration}\label{hh_calib}

Rather than carry out inverse modeling of the travel-time shifts to infer
the three-dimensional variation of the flows, we employ a simple calibration procedure. 
This minimizes complications
and uncertainties, due to the presence of strong photospheric magnetic fields,
in the inversion methods \citep[e.g.][]{DeGrave2014a}. Travel-time shifts
are related to the actual flows through a
convolution of the true flow components and the appropriate sensitivity
functions. For our measurements with a focus depth of 3 Mm, the 
sensitivity functions are sufficiently
compact in volume so as to render the 
travel-time differences 
reasonable proxies for the horizontal components of the near-surface 
flows themselves.
HH analyses of near-surface flows using travel-time shifts as 
flow proxies have been carried out
in prior studies \citep[e.g.][]{Braun2004,Braun2011a,Birch2013}.
Here, the travel-time differences $\tau_{we}$
and $\tau_{ns}$ are 
calibrated into westward and northward vector components 
of a horizontal depth-independent 
flow ($u_x$, $u_y$) by applying two different tracking rates 
to the same region of the Sun. These rates consist of (1) the nominal 
Carrington rotation rate
and (2) the nominal Carrington rate plus a constant offset. 
The tracking offset divided by the shift in $\tau_{we}$
between the sets of measurements 
yields a calibration factor of -7.5 
relating the speed (in units of m s$^{-1}$) to the travel-time 
difference (in s).
The minus sign reflects the fact that a 
positive flow directed
to the north (west) produces a reduction in the corresponding north-south
(west-east) time difference.
The range over the depth to which our calibrated flow is sensitive
is discussed in \cite{Braun2007}, which shows the sensitivity function
computed under the Born approximation,
for lateral-vantage HH at a 3 Mm focus-depth. 
The function has a broad contribution which is peaked around 3 Mm depth. Approximately
60\% of the sensitivity occurs between depths of 2--5 Mm, with
a 30\% (10\%) sensitivity for shallower (deeper) flows. 

\section{Flow Parameters}\label{flow_params}

Motivated by the 
the results of prior helioseismic surveys or analyses of
specific active-region flows,
we consider several flow parameters for study. 
These include the vertical component of the 
vorticity 
$$\textrm{VOR} = \frac{\partial u_y}{\partial x} -  \frac{\partial u_x}{\partial y},$$
and the horizontal divergence 
$$\textrm{DIV} = \frac{\partial u_y}{\partial x} +  \frac{\partial u_x}{\partial y}.$$
We also include a proxy for the vertical contribution to the kinetic
helicity \citep[][]{Rudiger1999,Komm2007} 
$$\textrm{HEL} = \textrm{VOR}\cdot \textrm{DIV}, $$
and the horizontal speed
$$|V| =  \sqrt{{u_x}^2 + {u_y}^2}.$$
The proxy $\textrm{HEL}$ is related to the vertical component of the true
kinetic helicity by a substitution of the horizontal divergence 
for the vertical component of the flow, with the ratio of the
two given by the density scale height
in the anelastic approximation \citep{Rudiger1999}.

If $x$ and $y$ denote westward and northward directions respectively,
than the sign of $\textrm{VOR}$ as defined above is positive (negative) for
counterclockwise (clockwise) vortical motion as viewed from above
the solar photosphere. Prior
analyses have shown that the vertical vorticity and kinetic helicity, with
this sign convention, have 
antisymmetric properties, at least statistically, with respect to
the equator.
This is true for convective
motions in both the quiet-Sun (supergranulation) field \citep{Duvall2000} and for motions
within active regions \citep{Komm2007}. 
To facilitate the combination of flow quantities
measured in ARs from both hemispheres, we 
switch the sign of $\textrm{VOR}$ in the southern hemisphere, so that
over the entire Sun
a positive (negative) value of $\textrm{VOR}$ is indicative of cyclonic (anticyclonic) motions.
A positive $\textrm{VOR}$, for example, implies counterclockwise rotation 
in the northern hemisphere and clockwise rotation in the southern hemisphere.

\subsection{Example: AR 11263}\label{ar11263}

\begin{figure}[htbp]
\epsscale{1.2}
\plotone{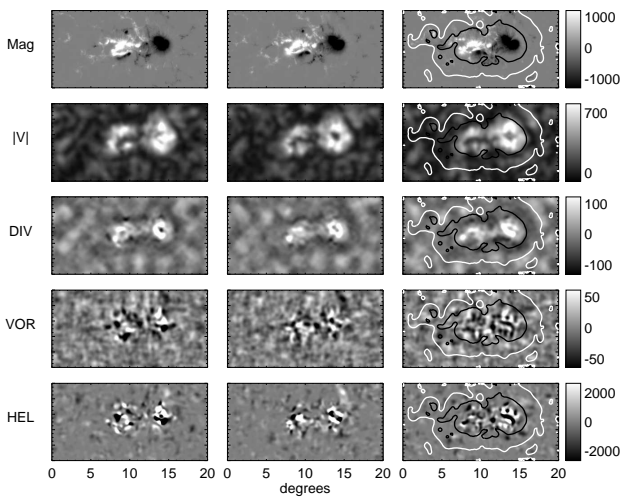}
\caption{
Magnetograms and flow parameter maps for AR 11263 for three successive 
time intervals.
From the top row down are 
maps of the: line-of-sight HMI magnetograms, near-surface
horizontal speed $|V|$, divergence ($\textrm{DIV}$) proxy, vorticity ($\textrm{VOR}$) proxy
and kinetic helicity ($\textrm{HEL}$) proxy (as defined in the text). The three
columns indicate successive 13.6 hr non-overlapping intervals over which the
quantities shown are obtained. The middle column represents observations
centered at 2011 Sept 4 00:58 UT. The contours in the right-most
column define magnetic-field flux densities of 50 (white) and 200 G 
(black) derived using a potential-field extrapolation and smoothed 
for the purpose of this figure. The grayscales cover the range of
values indicated, with scales in units of G for the magnetograms, m s$^{-1}$
for the speed $|V|$, $10^{-6}$ s$^{-1}$ for $\textrm{DIV}$ and $\textrm{VOR}$, and
$10^{-12}$ s$^{-2}$ for the helicity proxy $\textrm{HEL}$.
}
\label{fig.maps_11263}
\end{figure}

Figure~\ref{fig.maps_11263} shows maps of 
the four basic flow parameters, compared to time-averaged line-of-sight magnetograms,
for three consecutive time intervals and centered
on AR 11263. The derivatives for evaluating $\textrm{VOR}$ and $\textrm{DIV}$ are 
computed in the Fourier (horizontal wavevector) domain. To minimize
high-frequency noise all of the maps are smeared with a two-dimensional 
Gaussian with a FWHM of 0.48$^\circ$ (5.8Mm).
This region reached a maximum size of 720 $\mu$H (larger than 90\% of the 
sample) and was responsible for one X-class flare and 3 M-class flares. 
Qualitatively the features visible in the flow-parameter maps of AR 11263 
are fairly 
typical for our sample. The largest speeds, reaching nearly 1 km s$^{-1}$,
are found surrounding sunspots and are associated with the moat flows.
These moats also produce the largest diverging signals in the $\textrm{DIV}$ maps.
Surrounding these outflows in the $\textrm{DIV}$ maps are regions of converging flows
which extend, at most, a few degrees beyond the moats. The vorticity and
helicity maps exhibit complex patterns composed of compact features with
both signs and sizes as small as the resolution imposed by our 
0.48$^\circ$ smearing. There is a clear enhancement of the vorticity
(and helicity) signals within the strongest magnetic regions. 
Although some fraction of these signals may
be attributed to the effects of realization noise, a careful examination
of the maps reveals features which persist from one 14 hr interval to
the next. Successive vorticity maps, masked to isolate only
regions with flux densities greater than 500 G, are correlated with 
values of Pearson's coefficient equal to about 0.4. Successive divergence
maps have typical Pearson's coefficients of about 0.6. 

Supergranulation in the magnetic-free regions have peak speeds
on the order of $\approx$ 300 m s$^{-1}$ which appear as 
ring-like features (weaker than the sunspot moats) in the maps of $|V|$ and 
diverging centers surrounding by converging lanes in the $\textrm{DIV}$ maps. 

\begin{figure}[htbp]
\epsscale{1.2}
\plotone{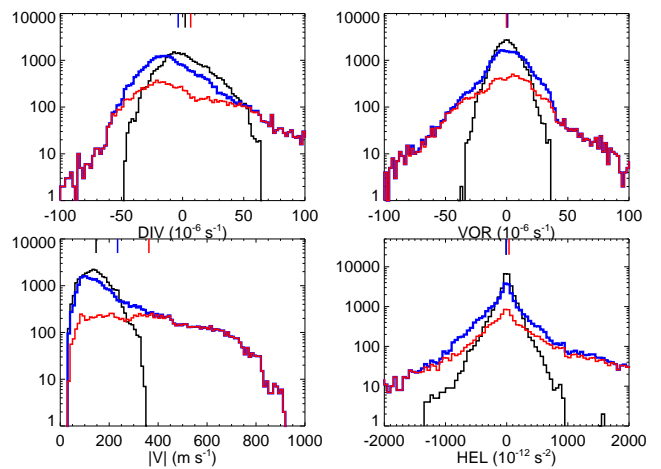}
\caption{
Distributions of the four flow parameters, $\textrm{DIV}$, $\textrm{VOR}$, $|V|$, and $\textrm{HEL}$ 
for one interval centered on AR 11263. The colors indicate
histograms of each flow parameter over different masks defined
by the (potential-field) magnetic flux-density  $B_p$:
$|B_p| > 200 $G (red), $|B_p| > 50 $G (blue) and $|B_p| < 50 $G (black).
The mean values of each distribution are indicated by the vertical markers, with
appropriate color, descending from the top of each panel. 
Note that in the wings of the distributions, the red and blue curves are 
nearly the same.
}\label{fig.pixel_dist}
\end{figure}

Our aim is to quantitatively condense the spatially complex maps, shown
for example in Figure~\ref{fig.maps_11263}, into manageable
``flow indices,'' which represent the the first
and second moments of their distribution over regions isolated by masks.
It is the magnetic field which must, by most conceivable means,
link any near-surface flow with the flaring process. 
Consequently, the purpose of the mask is to isolate the flows in magnetically relevant pixels
for further analysis. Recognizing that what defines ``relevant'' is 
unknown, we select different thresholds of photospheric magnetic field to
construct these masks. Specifically, two sets of masks
with minimum flux-density thresholds set at 50 and 200 G are used. The masks provide
alternately a fairly generous or restrictive definition of a magnetic region, with the goal
to establish whether and how the results we obtain depend on this choice.
We note that the stronger threshold primarily isolates the sunspots and their immediate
vicinities while the weaker value includes considerable surrounding field.
Figure~\ref{fig.pixel_dist} shows pixel histograms for our flow-parameter maps
for one interval centered on AR 11263 using the different masks.

A strict use of the (unsigned) 
line-of-sight magnetograms can produce highly discontinuous masks due
to, for example, false neutral lines in sunspot penumbrae which appear when
the spots are near the limb. Consequently, we use thresholds based
on maps of a potential-field extrapolation of the total field, $B_p$, from the line-of-sight
magnetograms. All of the masks are limited in the spatial domain by
a ``bounding box'' spanning 20$^\circ$ in longitude and 10$^\circ$ in latitude
and centered on each NOAA AR coordinate 
(e.g. the area shown in Figure~\ref{fig.maps_11263}).
This box was adequate for most of the active regions in our sample, however
five regions (ARs 11339, 11520, 11944, 11967, and 12192)
extend well beyond this area.  
Consequently a larger $(20^\circ \times 20^\circ)$ bounding box was used for 
these five regions. 
For the purpose of comparison, we also construct a ``quiet mask'' which
consists of all pixels within the bounding box with $|B_p| < 50$G.
Distributions for the flow parameters using these masks are indicated by
different colored histograms in 
Figure~\ref{fig.pixel_dist}. Hereafter, we refer to the three masks as
the ``200+ G mask,'' ``50+ G mask,'' and ``quiet mask'' respectively.

The distributions in Figure~\ref{fig.pixel_dist} are substantially wider for 
magnetic pixels than for the quiet mask, confirming the impression obtained by
a visual examination of the maps (e.g. Figure~\ref{fig.maps_11263}). It is also 
evident that the largest values of all the flow parameters are confined within the 200+ G
mask, 
which causes the wings of the histograms for the 200+ and 50+ G cases to coincide
nearly identically.
While $\textrm{VOR}$ and $\textrm{HEL}$ distributions are nearly symmetric
about zero, the distributions of the divergence parameter $\textrm{DIV}$ are skewed for all
masks, even for the quiet pixels. Prior studies of supergranular flows 
have also demonstrated asymmetric distributions of the horizontal 
divergence \citep[e.g.][]{Duvall2000}. It is also evident that, for the example
interval shown in Figure~\ref{fig.pixel_dist}, the $\textrm{DIV}$ parameter has a mean value
which is positive (diverging flows) for the 200+ G mask, and negative (converging
flows) for the mask constructed with the lower 50 G threshold. The former value is 
consistent with Figure~\ref{fig.maps_11263} where it can be seen that 
the 200+ G mask fully contains the diverging sunspot moat. The 50+ G mask
includes both the sunspot moats and apparently a sufficient area of
converging flow surround the moats to outweigh their contribution in the spatial average.

In studies of flows averaged over substantially larger areas, such as with low-resolution
RD methods, large-scale motions such as differential rotation are often removed from the data 
in order to isolate weaker AR-related flows \citep[e.g.][]{Komm2004,Komm2007}.
This is neither necessary nor desired 
with the strong compact flow signals revealed with
high-resolution HH. For reference, 
at a typical AR latitude of 15$^\circ$, the differential rotation contributes
a background vorticity of about 0.3 $\times 10^{-6}$ s$^{-1}$. This is two orders
of magnitude below 
the largest signals observed within the AR masks as 
shown in Figures~\ref{fig.maps_11263} 
and \ref{fig.pixel_dist}, and at any rate varies little across the dimensions of the magnetic masks. 

\subsection{Sample Distributions}\label{sample_distribs}

We consider for further analysis the set of 16 flow indices which
consist of the mean and standard deviations of the four basic flow parameters 
within either of the two flux-density masks. We use a 
notation such that, e.g. for the $\textrm{DIV}$ parameter, the quantities 
$\langle \textrm{DIV} \rangle_{50+}$ and $\std(\textrm{DIV})_{50+}$
represent the mean and standard deviation of $\textrm{DIV}$ over the pixels in
the 50+ G mask, and
$\langle \textrm{DIV} \rangle_{200+}$ and $\std(\textrm{DIV})_{200+}$
represent the analogous indices over the 200+ G mask.
This is also applied for the other 12 indices computed from the three
parameters $\textrm{VOR}$, $\textrm{HEL}$, and $|V|$ and the two masks. 
An additional 8 indices, representing the
mean and standard deviations of the four flow basic parameters, over the quiet pixels
within the bounding box, are also computed for purposes of comparison. These 
indices are denoted with a subscript $q$: for example, $\langle \textrm{DIV} \rangle_{q}$ and $\std(\textrm{DIV})_{q}$.

\begin{figure}[htbp]
\epsscale{1.2}
\plotone{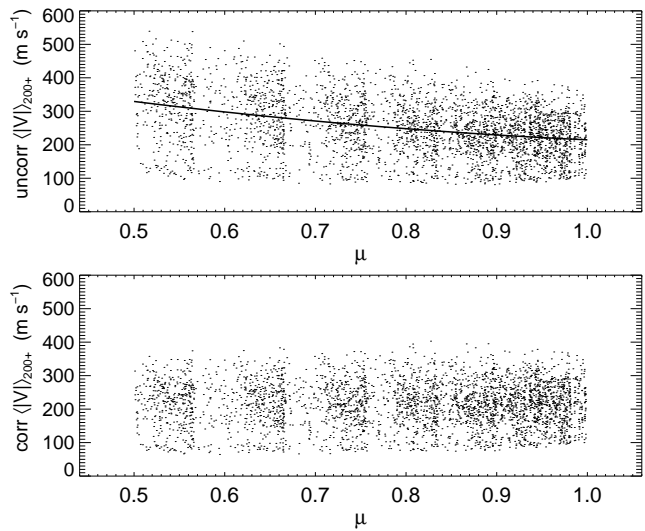}
\caption{
Top panel: uncorrected values of the speed $|V|$, averaged over magnetic pixels
with $|B_p| > 200 $G, as
a function of $\mu$, the cosine of the heliocentric angle. The solid line indicates
a quadratic polynomial fit $f$ to the data. 
Bottom panel: corrected values of $|V|$ after dividing out the normalized
function  $f(\mu)/f(1)$.
}\label{fig.mucorr}
\end{figure}

The indices $\langle \textrm{DIV} \rangle$, $\langle \textrm{VOR} \rangle$, and
$\langle \textrm{HEL} \rangle$ are signed quantities, while 
$\langle|V|\rangle$,  $\std(\textrm{DIV})$, $\std(\textrm{VOR})$, $\std(\textrm{HEL})$, and $\std(|V|)$ 
are not. We find a 
systematic variation in the unsigned quantities with distance of the AR from
the center of the disk. For example, Figure~\ref{fig.mucorr} illustrates this systematic variation
for the index $\langle |V| \rangle_{200+}$. Errors due to foreshortening may introduce
noise for the helioseismic measurements performed away from the center of the
disk. This noise appears be enhanced in ARs and is possibly related to the suppression of
wave amplitudes in magnetic areas. The result is that
the signed indices show spurious temporal variations as the AR rotates
across the disk, with higher (lower) values near the limbs (disk center). 
To remove this variation from each of the signed flow indices, a correction
factor $f(\mu)/f(1)$ is divided out of the uncorrected measurements, where $f$ is a 
quadratic fit to the data (e.g. Figure~\ref{fig.mucorr}). 
No corrections are applied to the signed indices for which trends with distance
from the central meridian (from east limb to west limb),
or with heliocentric angle, appear to be negligible (i.e.\ produce trend-related 
deviations on the order of a few percent or less of their standard deviation).
Artifacts, which appear as pseudo flows pointing towards or away from disk center,
are well known in other helioseismic measurements 
\citep[e.g.][]{Duvall2009,Zhao2012b,Baldner2012}. We note that
these large-scale low-amplitude ($\leq 10$ m s$^{-1}$) artifacts are very small
compared to the flows observed within our spatially compact (AR-sized) masks.

\begin{figure}[htbp]
\epsscale{1.2}
\plotone{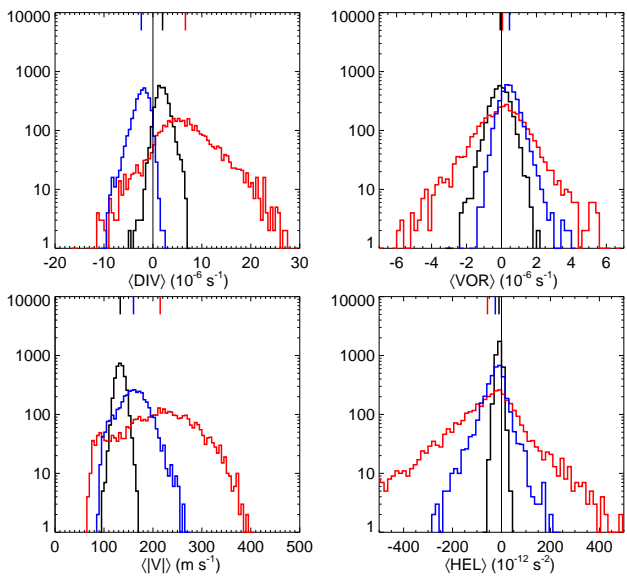}
\caption{
The distributions over the sample of measurements of several of the spatially-averaged
flow parameters. The red, blue, and black histograms represent spatial averages over
the 200+ G, 50+ G, and quiet masks respectively (see text).
The mean of each distribution
is shown by the vertical marker, with the appropriate color, descending from the top
of each panel.
}\label{fig.sample_dist}
\end{figure}

Figure~\ref{fig.sample_dist} presents histograms of eight of the flow indices for the entire
sample of measurements. We also include 
for comparison the distributions of the same four
parameters determined using the quiet-pixel masks. Not shown (for brevity) are the distributions
of the indices based on the standard deviations of the four basic parameters. Qualitatively,
they resemble the distributions of the unsigned indices based on 
averages of $\langle |V| \rangle$, which are
shown in the lower-left panel of Figure~\ref{fig.sample_dist}. 
In Figure~\ref{fig.sample_dist}, about 280 
(or 7\% of the total) sets of measurements 
are excluded from the histograms for which the number of pixels in the 
50+ G mask fell below $10^4$, which is about one-third of the
median area over the sample. This cut-off effectively removes
measurements centered on an active region during time intervals prior to
its eventual emergence in the photosphere.

There are a number of items worth
noting in this figure. The broadest distributions are observed for indices 
derived with the 200+ G mask. Thus, the strongest variation in the flow indices, among different
ARs and time intervals, occurs in the strongest magnetic regions. In addition, the spatially
averaged $\textrm{DIV}$ indices have clear sign preferences, depending on the magnetic mask.
Averaged over the 50+ G mask, the divergence is overwhelmingly negative, while the opposite
is true for the 200+ G mask. The nearby quiet regions have a distribution in 
$\langle \textrm{DIV} \rangle$ which is largely positive. It is plausible that this arises in large
measure from the presence of flows directed into the ARs which originate in the nearby 
quiet pixels and cross over the boundary defined by the 50 G threshold.
The averaged vorticity index for the 50+ G mask also shows a clear (cyclonic) sign preference,
but the quiet pixels and those above the 200 G threshold 
show a more symmetric distribution about zero.
Nevertheless, there is a clear preference for negative helicity as averaged
over all (even quiet) masks.  This arises if the spatially averaged values 
of $\textrm{DIV}$ and $\textrm{VOR}$ are anti-correlated,
as one might expect from the action of the Coriolis force due to solar rotation.

\subsection{Flare Productivity}\label{flow_vs_flares}

\begin{figure}[htbp]
\epsscale{1.2}
\plotone{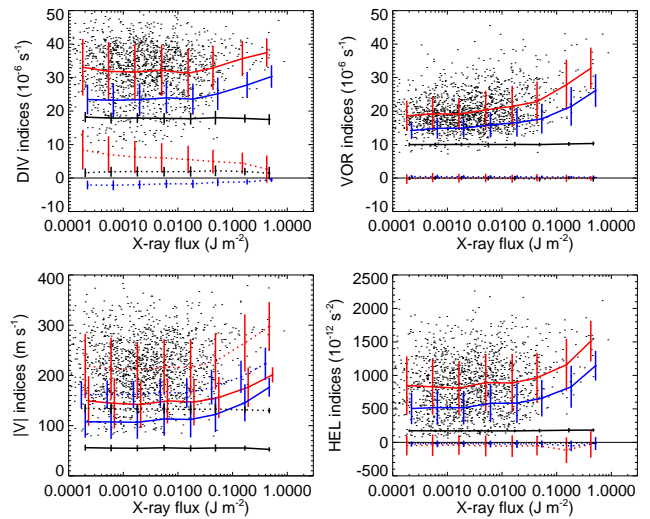}
\caption{
Flow indices as functions of the flare productivity, defined as the 
integrated X-ray flux summed over all flares within each AR and 13.6 hr time interval
included in our survey. 
The dotted (solid) lines connect flux-binned averages of indices representing the
mean (standard deviation) of the flow parameters indicated on the vertical axes. 
The colors
indicate the mask used, where red, blue and black curves indicate the 200+ G, 50+ G, and quiet
masks. The vertical bars indicate the range defined by the average 
$\pm$ one standard deviation for all points
within bins of the (logarithm of the) X-ray flux. 
For clarity, the scattered points show only the
individual measurements which went into the averages indicated by the top-most 
curve in each panel.
The vorticity and helicity indices based on the mask averages (dotted curves)
are difficult to discern in the two right panels and are replotted in
Figure~\ref{fig.close_up} with an expanded vertical scale.
}\label{fig.xray_flux}
\end{figure}

\begin{figure}[htbp]
\epsscale{1.2}
\plotone{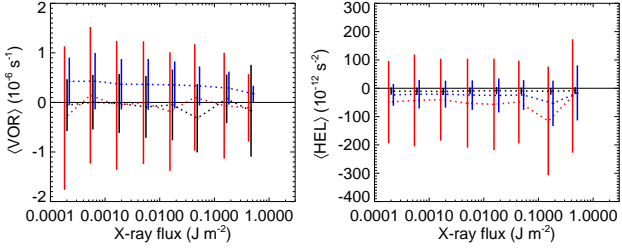}
\caption{
The indices based on the spatial averages of the 
vorticity (left panel) and helicity (right panel) 
as functions of the X-ray flux. 
These are the same quantities as shown in the right panels of 
Figure~\ref{fig.xray_flux} but with expanded vertical scales.
The red, blue, and black curves indicate the results for 
the 200+ G mask, 50+ G mask, and the quiet pixels, respectively.
The vertical bars indicate the range defined by the average 
$\pm$ one standard deviation for all points
within bins of the (logarithm of the) X-ray flux.
}\label{fig.close_up} 
\end{figure}

Before exploring the time variation of the flow indices, we 
examine the overall correlation between the indices 
with flare productivity over all ARs and time intervals.
We use NOAA event reports\footnote{\url{http:ngdc.noaa.gov/stp/space-weather/solar-data/solar-features/solar-flares/x-rays/goes/}}
which record the integrated X-ray flux (in J m$^{-2}$), times of the peak X-ray flux, 
and associated active region number, 
for solar flares as observed with the GOES spacecraft.
For each AR and each 13.6 hr time interval, 
we integrate the X-ray flux of any observed flares in that AR and time 
to produce a measure of the flare productivity. 
We first examine scatter plots of each of our flow indices against the
this flare productivity. For brevity,
Figures~\ref{fig.xray_flux} and \ref{fig.close_up} condenses the correlations indicated by these scatter
plots into binned averages of the 24 indices (i.e. the 16 AR indices and 8 quiet indices)
as functions of the X-ray flux. In general, the signed indices for the 50+ and 200+ G masks
and their quiet-mask counterparts exhibit little variation with flare productivity.
However, the the mean divergence assessed over the 200+ mask shows 
an apparent decrease with increasing X-ray flux. 
Further examination reveals a similar (decreasing towards zero) 
trend of the mean divergence with other indicators of the size of the active region, such as 
the integrated magnetic flux, or simply the number of pixels, in the 200+ G mask. 
The exact reason for this is unknown, but seems to indicate an increasing contribution of 
converging flows to the 200+ G mask as the size of active regions increases, even while
the mean divergence over the larger 50+ G mask stays nearly constant. 

\begin{figure}[htbp]
\epsscale{1.2}
\plotone{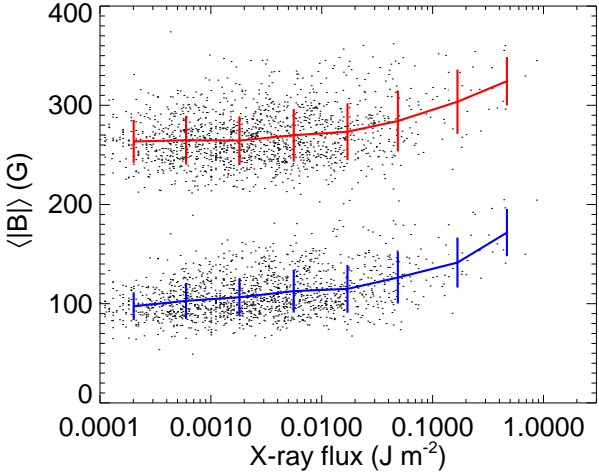}
\caption{
Scatter plots, with binned averages, of the unsigned magnetic flux density averaged over the 
200+ G (red) and 50+G (blue) masks respectively, as functions of the integrated flare X-ray flux. 
}\label{fig.mag_flare}
\end{figure}

The unsigned indices
$\langle|V|\rangle$,  $\std(\textrm{DIV})$, $\std(\textrm{VOR})$, $\std(\textrm{HEL})$, and $\std(|V|)$ 
over both magnetic masks
show increases for the most flare-active intervals (i.e. when
the X-ray flux exceeds about 0.1 J m$^{-2}$). The amount
of this increase is modest when compared to the spread of the measurements. Moreover,
the correlation is similar to that observed with other magnetic properties.
Figure~\ref{fig.mag_flare} shows that the averaged unsigned magnetic flux density,
over either the 50+ or 200+ G masks, has a similar dependence with X-ray flux. 
The similarity of the correlations between Figures \ref{fig.xray_flux} 
and \ref{fig.mag_flare} raises the question of whether there exists a physical 
link between the observed flow parameters and the onset of flares, or rather
simply that larger ARs provide a preferred environment for both increased flows and 
(as is already known) flare occurrence.

\section{Time Variation of Flow Parameters}\label{time_variations}

\begin{figure}[htbp]
\epsscale{1.2}
\plotone{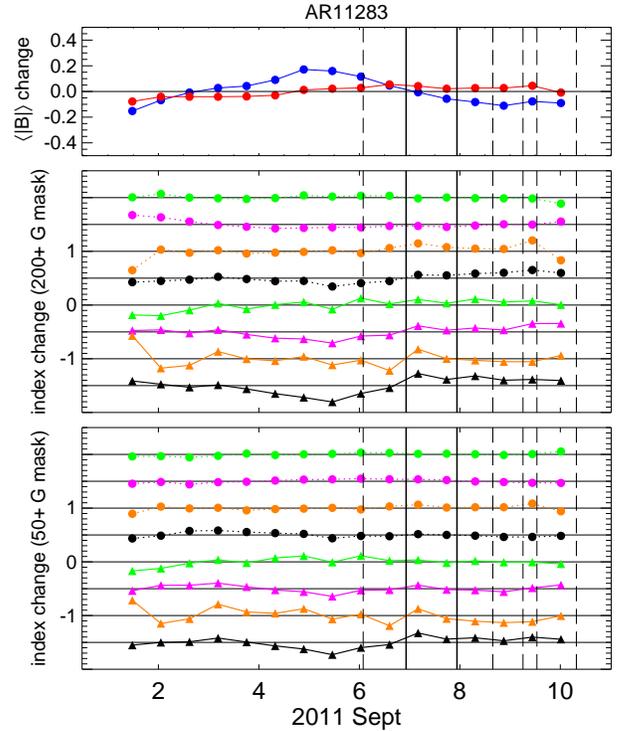}
\caption{
The variation in time of the mean unsigned magnetic flux density (top panel) 
and flow indices determined for the 200+ G mask (middle panel) and
50+ mask (bottom panel) G for AR 11283. All quantities are expressed
in terms of their fractional change from the time-averaged quantity
(see text for details). In the top panel, the red (blue) circles
connected by lines of the same color, indicate the fractional change
of the unsigned flux density averaged over the 200+ G (50+ G) masks.
In the middle and bottom panels, the quantities shown are (from top to
bottom) the fractional changes of: 
$\langle \textrm{VOR} \rangle$ (green circles),
$\langle \textrm{DIV} \rangle$ (purple circles),
$\langle \textrm{HEL} \rangle$ (orange circles),
$\langle |V| \rangle$ (black circles),
$\std(\textrm{VOR})$ (green triangles),
$\std(\textrm{DIV})$ (purple triangles),
$\std(\textrm{HEL})$ (orange triangles), and
$\std(|V|)$ (black triangles). 
For clarity,
each plot in the lower two panels 
is vertically displaced from the origin by multiples 
of 0.5 and denoted by the horizontal black lines.
Solid (dashed) vertical lines denote the times of X-class (M-class)
flares associated with this region.
}\label{fig.11283}
\end{figure}

\begin{figure}[htbp]
\epsscale{1.2}
\plotone{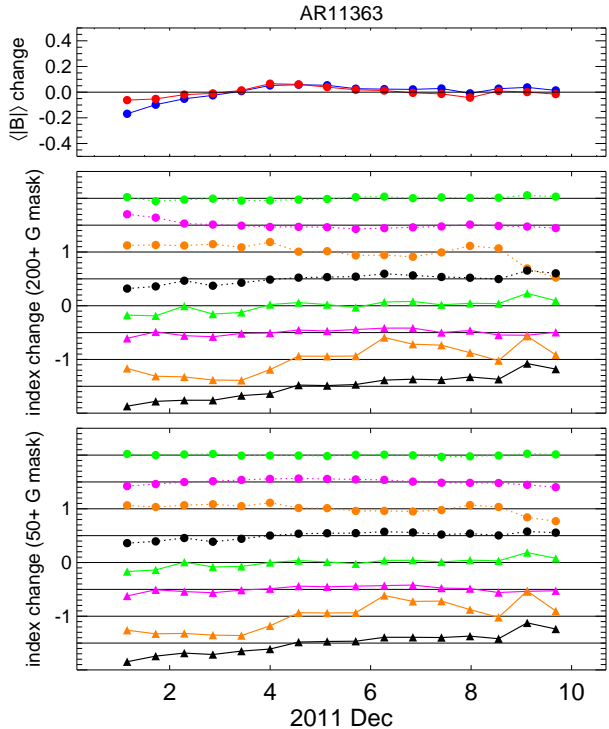}
\caption{
The variation in time of the unsigned magnetic flux density 
flow indices for AR 11363, shown in the same format as Figure~\ref{fig.11283}.
There are no X-class or M-class flares associated with this region
during the dates observed.
}\label{fig.11363}
\end{figure}

Figures~\ref{fig.11283} and \ref{fig.11363} show examples 
of the temporal variations of the flow indices for two ARs. 
One of these regions, AR 11283 
was the site of one X-class flare and 3 M-class flares, while the other, AR 11363
was a large, but relatively flare-free region (producing only 8 C-class 
flares during its passage).
The indices are plotted in terms of their fractional deviation from a
temporal mean. For the unsigned indices, e.g. $\std(\textrm{VOR})_{50+}$, this 
takes a form 
$$
\Delta \std(\textrm{VOR})_{50+} \equiv \frac{\std(\textrm{VOR})_{50+}-\overline{\std(\textrm{VOR})}_{50+}}{{\overline{\std(\textrm{VOR})}}_{50+}},
$$
where the horizontal bar denotes the average over time. For the signed quantities,
the change is relative to the corresponding time-average of the 
standard-deviation based index, e.g.
$$
\Delta \langle \textrm{VOR} \rangle_{50+} \equiv \frac{\langle \textrm{VOR} \rangle_{50+}-\overline{\langle \textrm{VOR} \rangle_{50+}}}{\overline{\std(\textrm{VOR})}_{50+}},
$$
and likewise for the other signed indices.
For AR 11283, vertical lines denote the times of the X-class and M-class flares.
It is inevitable that flows, and thus the flow indices, 
will change as an AR evolves. 
The two regions shown in 
Figures~\ref{fig.11283} and \ref{fig.11363} highlight the
dilemma in identifying flare-related flow changes. AR 11363 has no large 
flares, but both regions undergo evolution of at least some of the flow indices, e.g.
$\Delta \std(\textrm{HEL})_{50+}$ or $\Delta \std(|V|)_{50+}$. In light of
this, it is not clear how to interpret the variation of, say, 
$\Delta \std(|V|)_{50+}$ immediately prior to the onset of the X-class flare
in AR 11283 on 2011 Sept 6. 

Superposed Epoch (SPE) analysis \citep[e.g.][]{Mason2010,Reinard2010}
of the flow indices 
provides one method to reduce the influence of quiescent (non-flare related) 
evolution while revealing possibly subtle flare-related changes.
To achieve this, we identify flare times from the NOAA GOES records
of all X and M-class flares associated with the ARs in our survey. Each set
of flow measurements are assigned to a 13.7 hr-long time bin, based on the midpoint
in time of the HMI Dopplergrams used in the analysis relative to the time of the
peak X-ray flux of the flare. The indices within each bin are averaged,
using 6 bins before and 6 bins after the
onset of flares. Thus, the analysis covers about 169 hr (6.8 days) of time
around the occurrence of the flares. A data-coverage requirement was strictly
enforced that, for any candidate flare, the relevant flow measurements are
included in the SPE averages if, and only if, there are no gaps in those measurements
over the entire 6.8 days. As discussed earlier, this may arise because
of either unfavorable AR position (greater than 60$^\circ$ from disk center) or
gaps in the HMI data. This requirement ensures an equal weighting, over the entire
6.8 days considered, of the contribution of each set of measurements. 
In spite of this restriction, we are left with 70 flares (associated
with 34 distinct ARs) in our binned averages. 
We note that all X and M flares within each AR are included in the SPE analysis 
as the HH measurements allow, without rejection due to their 
proximity in time to other flares in the same region.

\begin{figure}[htbp]
\epsscale{1.2}
\plotone{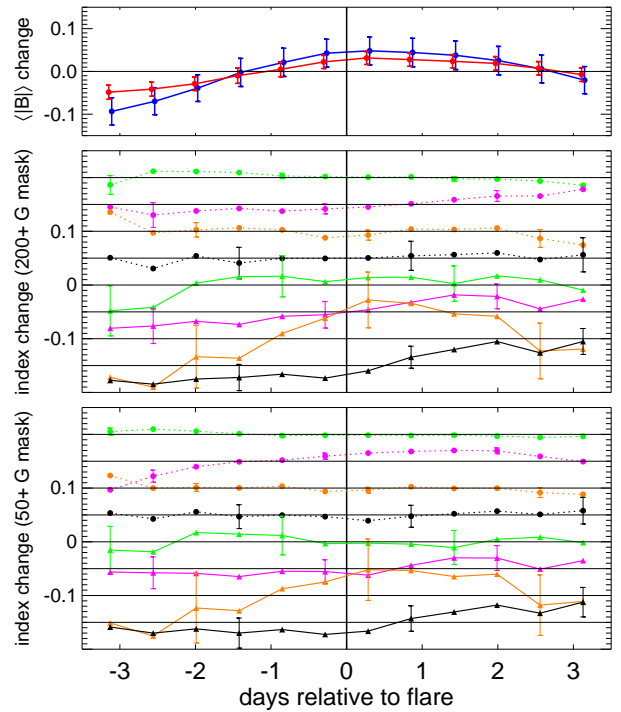}
\caption{
The variation in time of the SPE averaged flow indices over 70
X-class and M-class flares. The format is the same as for
Figure~\ref{fig.11363} except that the abscissa shows the time offset in days
from the midpoint of the bin intervals to the flare occurrence. 
Note that the vertical scale in all panels is substantially magnified 
compared to Figures~\ref{fig.11283}
and \ref{fig.11363} and that, in the lower two panels, 
offsets by multiples of 0.05 are applied to vertically separate
each plot for clarity.
Error bars represent the standard error of the mean.
To avoid clutter, only representative error bars are shown.
}\label{fig.avg6}
\end{figure}

Figure~\ref{fig.avg6} shows the results for the 70-flare average. Along 
with the flow indices, the figure also shows the SPE averages of the 
magnetic flux densities, determined
over both the 50+ G and 200+ G masks. The averaged flux density show significant 5-10\%
quadratic-like variations with time relative to the flares. It is most likely that
this results from foreshortening and projection effects related to the tendency of the 
AR to be positioned closer to the east (west) 
limb at early (later) times in the SPE averages than near the times of
the flares. Indeed, without the center-to-limb detrending performed on the 
unsigned flow variables (e.g. Figure~\ref{fig.mucorr}) one sees similar temporal
variations of those quantities as well.

It is apparent (particularly by noting the units of the vertical scales), that
the variations over time of the SPE averaged indices are considerably smaller
than observed within individual ARs.
For example, fractional variations of up to 40\% in $\std(\textrm{HEL})_{50+}$ 
or $\std(|V|)_{50+}$ 
are present in Figures~\ref{fig.11283} and \ref{fig.11363}.
The same quantities in the SPE-based averages show 
variations less than 10\%. Moreover, although the latter variations 
are still notable, they are not substantially larger than expected from the
standard error of the mean as indicated by the error bars. 
Most SPE-averaged flow indices, including the signed indices, exhibit 
time variations on the order of only a few percent over the entire 6.8 days. On the other hand,
some variations do appear significant when, for example, compared to their standard errors.
This includes variations in $\langle \textrm{DIV} \rangle_{50+}$, which show $\approx 3$\% 
temporal variations which are on the order of three times the standard errors.

To assess whether the observed variations in the SPE-averaged indices 
are associated with the
specific occurrence of flares, we perform a separate 
SPE averaging procedure substituting the GOES flare list with an equivalent list
with randomized times. In this control
list each true flare-time is offset by 
either $\delta$ or ($\delta - 13$) days, 
where $\delta$ is a random time between 0 and 
13 days, and the choice is limited to the offset 
which keeps the AR visible to HMI. 
Pre-emergence intervals are excluded from the SPE by
requiring a minimum pixel count of $10^4$ over the 50+ G mask 
during the interval of the (pseudo) flare.

\begin{figure}[htbp]
\epsscale{1.2}
\plotone{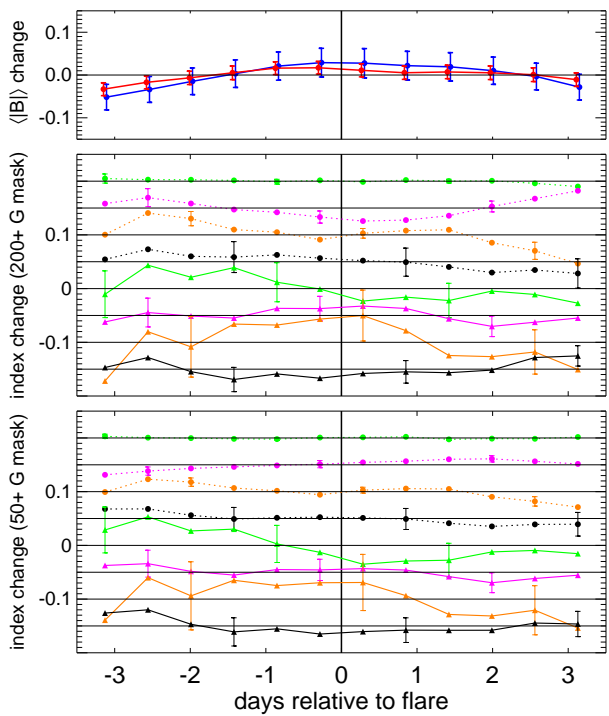}
\caption{
The variation in time of the SPE averages of flow indices with respect to a set
of 71 random moments in time, selected by applying random offsets from
the times of actual M and X-class flares.
}\label{fig.avg6rand}
\end{figure}

From this control list, we obtained the results shown in Figure~\ref{fig.avg6rand},
based on averages with respect to 71 pseudo flares.
Compared with Figure~\ref{fig.avg6}, the similarity is quite striking.
Just as in the SPE analysis with respect to actual X and M-class flare times,
there are small, but significant temporal variations
in several flow parameters. This is also apparently true for the magnetic flux density
as well. Indeed, the similarity in the time variation of $\langle |B| \rangle$ for the two cases
reinforces the likelihood of center-to-limb systematics, on the order of 5-10\%,
associated with the determination of the line-of-sight magnetic field.
Many of the the flow indices, such as $\langle \textrm{DIV} \rangle$ and $\std(\textrm{HEL})$, show
significant variations with respect to the random times, for one or both magnetic masks, 
which are similar in magnitude to those observed with respect to the true flare times.

We have performed SPE analysis over different flare classes (e.g. X-class
flares alone, or flares defined by their integrated, rather than peak, X-ray flux).
Decreasing the length of the time interval about flare occurrence increases
the number of available measurements and therefore the statistics. However, in all
cases examined, the results are qualitatively the same as illustrated with
the examples shown in Figures~\ref{fig.avg6}. 
We have also specifically examined the possibility of changes in the flow properties
at time-scales shorter than 13.6 hr. For a subset of ARs, specifically those responsible
for X-class flares, the HH analysis described in \S~\ref{hh} is repeated
using datacubes 4.5 hr in length (one third the length of the original analysis). 
For this subsample of ARs, we find no compelling evidence for significant changes in the
flow parameters on these shorter timescales.

\section{Discussion}\label{discussion}

We have carried out helioseismic survey of the near-surface flow parameters of 252 of the 
largest NOAA-numbered sunspot regions, spanning a 4.5 year interval between 2010 June and  
2014 December. Although other comprehensive helioseismic surveys have been made, for example,
with data from GONG, our survey includes flow measurements made with greater
spatial resolution.
The general properties of the horizontal flow divergence and vertical vorticity 
are similar to, and confirm, prior
results made at lower resolutions \citep[e.g.][]{Komm2004,Hindman2009}. Specifically,
we see that the typical active region is characterized by strong outflows by sunspots in the AR
centers, and surrounded by somewhat weaker converging counter cells. 
The vorticity signals are considerably weaker than those associated with the horizontal 
divergence, and likely include a larger fraction of realization noise. Nevertheless, active
regions show small-scale vorticity signals significantly higher than surrounding quiet regions.
The spatially-averaged kinetic helicity proxy for different magnetic masks indicate small rotational
motion consistent with Coriolis forces acting on the converging or diverging flows components.

There is a modest correlation with integrated flare X-ray flux of several of the flow 
indices we examined, 
including the average speed, and the standard deviations over the magnetic masks of all four 
basic parameters (speed, divergence, vorticity and kinetic helicity). However
these correlations strongly resemble those obtained with basic magnetic properties, such
as the mean unsigned flux density. 
The similarity of the results between the choice of the flux-density threshold in the
spatial masks shows that these correlations primarily arise for flows within or near
the strongest flux and are little changed when weaker flux regions are included.
Overall, the trends resemble at least qualitatively
the findings of \cite{Komm2009}, albeit with substantially different flow parameters.
We cannot at this stage rule out the
possibility that information characterizing the near-surface flows may, along with
magnetic indicators, help distinguish between flaring and non-flaring AR or even
predict flares such as suggested by \cite{Komm2011}. 

Some differences between our flow indices and those employed previously are worth noting.
The ``vorticity'' indicators defined by \cite{Komm2009} were 
constructed from relatively
large spatial variations of the horizontal components of vorticity, unlike the vertical
component examined in the present study. Furthermore, it is difficult to directly compare our near-surface
indices with the parameter defined by \cite{Reinard2010} which is proportional to, among other factors, the spread
in depth of kinetic helicity.
On the other hand, in light of these results, it is reasonable to question why higher
resolution observations of converging flows (e.g. such as characterized by the divergence-based indices
examined here) do not show stronger association with flaring regions than other indicators. 
This might be expected since converging flows presumably comprise the near-surface components of the ``vortex rings'' 
\citep{Komm2011b} inferred from the ring-diagram results to be strongly associated with flaring regions.

We do not see, in our own flare indices, specific events occurring near flares 
and which resemble the bumps in the vertical component of kinetic helicity 
suggested by \cite{Gao2014}. That study used direct inferences (from inverse
modeling of time-distance travel-time shifts) 
of vertical velocities within 1 Mm of the photosphere, but their kinetic helicity parameter is otherwise
directly comparable to our own proxy $\textrm{HEL}$. We note that 
\cite{Gao2014} used shorter, overlapping, 
time intervals (i.e. 8-hr long analyses spaced every 4 hr) but looked at a much
smaller sample of flaring regions than considered here.
\cite{Gao2014} also examined the 
so-called ``normalized helicity gradient variance'' of \cite{Reinard2010} with results
not substantially different from those obtained with the simpler helicity measurements.

The main result of our survey is that individual ARs, 
and averages made with superposed epoch analysis of ARs with respect to the times
of strong flares, show mostly variations associated with (non-flare related) evolution.
The similarity between the SPE results obtained for the two flux-density thresholds used
also implies that most of these variations are restricted to the stronger flux regions near
and within sunspots.
Indeed, it appears challenging to remove this intrinsic variation in order to detect
possible, specific, flare-related trends in the indices.
The SPE analysis reduces much, but not all, evolution-related changes. 
Thus, even with
nearly five years of high-quality data, it is apparently 
difficult to collect sufficient numbers of
strong flares which also occur favorably close to the center of the disk to 
accommodate several days of helioseismic processing of both pre-flare and post-flare 
data. As we have noted, we 
performed additional SPE analysis with increased numbers 
of flares by reducing the time window around the flares.
Our tests to date give essentially
similar results. Namely, there are no temporal variations of the flow indices which are greater than
observed in control tests with similar numbers of flares reassigned to random times.

Future work employing a wider parameter-search might be useful see whether flows with specific
properties, or occurring within more restricted locations than the spatial masks used here, 
are more strongly correlated with flares. Such parameters could include properties of the
flow field, the overlying magnetic field, and the vicinity (or connectivity of 
the field) to actual flare sites.  A careful examination of individual case studies 
with data from HMI and other instruments could serve to guide this parameter selection. 

\acknowledgments
We are grateful to Charles Baldner and the rest of
Helioseismic and Magnetic Imager (HMI) team at Stanford
University for computing and providing the custom datasets 
for this study. Many thanks to KD Leka for 
enlightening discussions and 
a careful read of a draft of the manuscript.  
We are also appreciative of the helpful suggestions made by the referee.
This work is supported by the Solar 
Terrestrial program of the National Science Foundation
through grant AGS-1127327 awarded to NWRA.
Additional support is provided by the NASA
Heliophysics program through contracts NNH12CF23C and NNH12CF68C. 
SDO data is provided courtesy of NASA/SDO and the AIA, EVE, and HMI
science teams.

\bibliographystyle{/export/home/dbraun/Macros/apj}
\bibliography{/export/home/dbraun/Macros/db.bib}

\begin{thebibliography}{37}
\expandafter\ifx\csname natexlab\endcsname\relax\def\natexlab#1{#1}\fi

\bibitem[{{Baldner} \& {Schou}(2012)}]{Baldner2012}
{Baldner}, C.~S. \& {Schou}, J. 2012, \apjl, 760, L1

\bibitem[{{Beauregard} {et~al.}(2012){Beauregard}, {Verma}, \&
  {Denker}}]{Beauregard2012}
{Beauregard}, L., {Verma}, M., \& {Denker}, C. 2012, Astronomische Nachrichten,
  333, 125

\bibitem[{{Birch} {et~al.}(2013){Birch}, {Braun}, {Leka}, {Barnes}, \&
  {Javornik}}]{Birch2013}
{Birch}, A.~C., {Braun}, D.~C., {Leka}, K.~D., {Barnes}, G., \& {Javornik}, B.
  2013, \apj, 762, 131

\bibitem[{{Braun}(2014)}]{Braun2014}
{Braun}, D.~C. 2014, \solphys, 289, 459

\bibitem[{{Braun} \& {Birch}(2008)}]{Braun2008b}
{Braun}, D.~C. \& {Birch}, A.~C. 2008, \apjl, 689, L161

\bibitem[{{Braun} {et~al.}(2007){Braun}, {Birch}, {Benson}, {Stein}, \&
  {Nordlund}}]{Braun2007}
{Braun}, D.~C., {Birch}, A.~C., {Benson}, D., {Stein}, R.~F., \& {Nordlund}, A.
  2007, \apj, 669, 1395

\bibitem[{{Braun} {et~al.}(2004){Braun}, {Birch}, \& {Lindsey}}]{Braun2004}
{Braun}, D.~C., {Birch}, A.~C., \& {Lindsey}, C. 2004, in SOHO 14 Helio- and
  Asteroseismology: Towards a Golden Future, ed. D.~{Danesy}, Vol. 559
  (Noordwijk: ESA), 337--340

\bibitem[{{Braun} \& {Wan}(2011)}]{Braun2011a}
{Braun}, D.~C. \& {Wan}, K. 2011, J. Phys.: Conf. Ser., 271, 012007

\bibitem[{{Brickhouse} \& {LaBonte}(1988)}]{Brickhouse1988}
{Brickhouse}, N.~S. \& {LaBonte}, B.~J. 1988, \solphys, 115, 43

\bibitem[{{DeGrave} {et~al.}(2014){DeGrave}, {Jackiewicz}, \&
  {Rempel}}]{DeGrave2014a}
{DeGrave}, K., {Jackiewicz}, J., \& {Rempel}, M. 2014, \apj, 788, 127

\bibitem[{{Duvall} \& {Gizon}(2000)}]{Duvall2000}
{Duvall}, Jr., T.~L. \& {Gizon}, L. 2000, \solphys, 192, 177

\bibitem[{{Duvall} \& {Hanasoge}(2009)}]{Duvall2009}
{Duvall}, Jr., T.~L. \& {Hanasoge}, S.~M. 2009, in Solar-Stellar Dynamos as
  Revealed by Helio- and Asteroseismology: GONG 2008/SOHO 21, ed. M.~{Dikpati},
  T.~{Arentoft}, I.~{Gonz{\'a}lez Hern{\'a}ndez}, C.~{Lindsey}, \& F.~{Hill},
  Vol. 416 (San Francisco: Astron. Soc. Pacific), 103--109

\bibitem[{{Fan}(2009)}]{Fan2009}
{Fan}, Y. 2009, Liv Rev Solar Phys, 1

\bibitem[{{Gao} {et~al.}(2014){Gao}, {Zhao}, \& {Zhang}}]{Gao2014}
{Gao}, Y., {Zhao}, J., \& {Zhang}, H. 2014, \solphys, 289, 493

\bibitem[{{Gizon} {et~al.}(2001){Gizon}, {Duvall}, \& {Larsen}}]{Gizon2001}
{Gizon}, L., {Duvall}, Jr., T.~L., \& {Larsen}, R.~M. 2001, in IAU Symposium,
  Vol. 203, Recent Insights into the Physics of the Sun and Heliosphere:
  Highlights from SOHO and Other Space Missions, ed. P.~{Brekke}, B.~{Fleck},
  \& J.~B. {Gurman}, 189--191

\bibitem[{{Haber} {et~al.}(2001){Haber}, {Hindman}, {Toomre}, {Bogart}, \&
  {Hill}}]{Haber2001}
{Haber}, D.~A., {Hindman}, B.~W., {Toomre}, J., {Bogart}, R.~S., \& {Hill}, F.
  2001, in SOHO 10/GONG 2000 Workshop: Helio- and Asteroseismology at the Dawn
  of the Millennium, ed. A.~{Wilson} \& P.~L. {Pall{\'e}}, Vol. 464 (ESA),
  209--212

\bibitem[{{Haber} {et~al.}(2004){Haber}, {Hindman}, {Toomre}, \&
  {Thompson}}]{Haber2004}
{Haber}, D.~A., {Hindman}, B.~W., {Toomre}, J., \& {Thompson}, M.~J. 2004,
  \solphys, 220, 371

\bibitem[{{Hindman} {et~al.}(2009){Hindman}, {Haber}, \&
  {Toomre}}]{Hindman2009}
{Hindman}, B.~W., {Haber}, D.~A., \& {Toomre}, J. 2009, \apj, 698, 1749

\bibitem[{{Komm} {et~al.}(2004){Komm}, {Corbard}, {Durney}, {Gonz{\'a}lez
  Hern{\'a}ndez}, {Hill}, {Howe}, \& {Toner}}]{Komm2004}
{Komm}, R., {Corbard}, T., {Durney}, B.~R., {Gonz{\'a}lez Hern{\'a}ndez}, I.,
  {Hill}, F., {Howe}, R., \& {Toner}, C. 2004, \apj, 605, 554

\bibitem[{{Komm} {et~al.}(2011{\natexlab{a}}){Komm}, {Ferguson}, {Hill},
  {Barnes}, \& {Leka}}]{Komm2011}
{Komm}, R., {Ferguson}, R., {Hill}, F., {Barnes}, G., \& {Leka}, K.~D.
  2011{\natexlab{a}}, \solphys, 268, 389

\bibitem[{{Komm} \& {Gosain}(2015)}]{Komm2015c}
{Komm}, R. \& {Gosain}, S. 2015, \apj, 798, 20

\bibitem[{{Komm} \& {Hill}(2009)}]{Komm2009}
{Komm}, R. \& {Hill}, F. 2009, Journal of Geophysical Research (Space Physics),
  114, 6105

\bibitem[{{Komm} {et~al.}(2011{\natexlab{b}}){Komm}, {Howe}, {Hill}, \&
  {Jain}}]{Komm2011b}
{Komm}, R., {Howe}, R., {Hill}, F., \& {Jain}, K. 2011{\natexlab{b}}, in IAU
  Symposium, Vol. 273, The Physics of Sun and Star Spots, ed. D.~{Prasad
  Choudhary} \& K.~G. {Strassmeier}, 148--152

\bibitem[{{Komm} {et~al.}(2007){Komm}, {Howe}, {Hill}, {Miesch}, {Haber}, \&
  {Hindman}}]{Komm2007}
{Komm}, R., {Howe}, R., {Hill}, F., {Miesch}, M., {Haber}, D., \& {Hindman}, B.
  2007, \apj, 667, 571

\bibitem[{{Leka} \& {Barnes}(2007)}]{Leka2007}
{Leka}, K.~D. \& {Barnes}, G. 2007, \apj, 656, 1173

\bibitem[{{Lindsey} \& {Braun}(1997)}]{Lindsey1997}
{Lindsey}, C. \& {Braun}, D.~C. 1997, \apj, 485, 895

\bibitem[{{Lindsey} \& {Braun}(2004)}]{Lindsey2004b}
---. 2004, \apjs, 155, 209

\bibitem[{{Mason} {et~al.}(2006){Mason}, {Komm}, {Hill}, {Howe}, {Haber}, \&
  {Hindman}}]{Mason2006}
{Mason}, D., {Komm}, R., {Hill}, F., {Howe}, R., {Haber}, D., \& {Hindman},
  B.~W. 2006, \apj, 645, 1543

\bibitem[{{Mason} \& {Hoeksema}(2010)}]{Mason2010}
{Mason}, J.~P. \& {Hoeksema}, J.~T. 2010, \apj, 723, 634

\bibitem[{{Reinard} {et~al.}(2010){Reinard}, {Henthorn}, {Komm}, \&
  {Hill}}]{Reinard2010}
{Reinard}, A.~A., {Henthorn}, J., {Komm}, R., \& {Hill}, F. 2010, \apjl, 710,
  L121

\bibitem[{{R{\"u}diger} {et~al.}(1999){R{\"u}diger}, {Brandenburg}, \&
  {Pipin}}]{Rudiger1999}
{R{\"u}diger}, G., {Brandenburg}, A., \& {Pipin}, V.~V. 1999, Astronom. Nach.,
  320, 135

\bibitem[{{Schrijver}(2007)}]{Schrijver2007}
{Schrijver}, C.~J. 2007, \apjl, 655, L117

\bibitem[{{Schrijver}(2009)}]{Schrijver2009}
---. 2009, Advances in Space Research, 43, 739

\bibitem[{{Wang} {et~al.}(2014){Wang}, {Liu}, {Deng}, \& {Wang}}]{Wang2014}
{Wang}, S., {Liu}, C., {Deng}, N., \& {Wang}, H. 2014, \apjl, 782, L31

\bibitem[{{Wang} {et~al.}(2011){Wang}, {Liu}, \& {Wang}}]{Wang2011}
{Wang}, S., {Liu}, C., \& {Wang}, H. 2011, in IAU Symposium, Vol. 273, The
  Physics of Sun and Star Spots, ed. D.~{Prasad Choudhary} \& K.~G.
  {Strassmeier}, 412--416

\bibitem[{{Welsch} {et~al.}(2009){Welsch}, {Li}, {Schuck}, \&
  {Fisher}}]{Welsch2009}
{Welsch}, B.~T., {Li}, Y., {Schuck}, P.~W., \& {Fisher}, G.~H. 2009, \apj, 705,
  821

\bibitem[{{Zhao} {et~al.}(2012){Zhao}, {Nagashima}, {Bogart}, {Kosovichev}, \&
  {Duvall}}]{Zhao2012b}
{Zhao}, J., {Nagashima}, K., {Bogart}, R.~S., {Kosovichev}, A.~G., \& {Duvall},
  Jr., T.~L. 2012, \apjl, 749, L5

\end{thebibliography}

\end{document}